\documentstyle[aps,multicol,epsf]{revtex}
\begin{document}
\draft
\title{Dynamics of Surface Roughening with Quenched Disorder}

\author{S. Havlin,$^{1,2}$ L.~A.~N. Amaral,$^1$ S.~V. Buldyrev,$^1$
S.~T. Harrington,$^1$ and H.~E. Stanley$^1$}

\address{$^1$Center for Polymer Studies and Dept. of Physics,
 Boston University, Boston, MA 02215 USA \\ $^2$Minerva Center and
Dept. of Physics, Bar-Ilan University, Ramat Gan, Israel }

\date{\today}

\maketitle

\begin{abstract}

  We study the dynamical exponent $z$ for the directed percolation
depinning (DPD) class of models for surface roughening in the presence
of quenched disorder.  We argue that $z$ for $(d+1)$ dimensions is
equal to the exponent $d_{\rm min}$ characterizing the shortest path
between two sites in an isotropic percolation cluster in $d$
dimensions.  To test the argument, we perform simulations and
calculate $z$ for DPD, and $d_{\rm min}$ for percolation, from $d = 1$
to $d = 6$.

\end{abstract}


\begin{multicols}{2}

  Recently the growth of rough interfaces has been the object of many
theoretical, numerical, and experimental studies, fueled by the broad
interdisciplinary aspects of the subject \cite{review}.  Applications
are as diverse as imbibition in porous media, fluid--fluid
displacement, bacterial colony growth, fire front motion, and the
motion of flux lines in superconductors \cite{review}.

  In general, a $d$-dimensional self-affine interface, described by a
single-valued function $h({\bf x},t)$, evolves in a
$(d+1)$-dimensional medium.  Some form of disorder, $\eta$, affects
the motion of the interface leading to its roughening.  Two main
classes of disorder have been discussed in the literature.  The first,
called ``annealed'', depends only on time.  The second, ``quenched''
disorder, is frozen in the medium.

  Continuum equations, such as the Kardar-Parisi-Zhang (KPZ) equation
\cite{kpz}, have been remarkably successful in describing roughening
for the case of annealed disorder \cite{review}.  For the quenched
disorder case, several models were proposed with a view toward
explaining experimental results for which the roughness exponent
$\alpha$ is significantly larger than the predictions for annealed
disorder (for reviews, see e.g. \cite{review}).  Here $W \sim
L^{\alpha}$, where $W$ is the interface width and $L$ is the system
size.

  For one class of models, the static properties of the interface in
$(1+1)$ dimensions are obtained exactly by a mapping, at the depinning
transition, onto {\it directed percolation\/} (DP) \cite{Havlin,dpd}.
In higher dimensions the mapping is to {\it directed surfaces\/} (DS)
\cite{Buldyrev} --- for $(1+1)$ dimensions, DP and DS are equivalent.
This class of models is referred to as {\it directed percolation
depinning\/} (DPD) \cite{alpha}.

  A recent numerical study \cite{Amaral+Makse}, confirmed by analytical
results \cite{Kardar}, showed that the DPD class of models can be
described by a stochastic differential equation of the KPZ type with
quenched disorder
\begin{equation}
\frac{\partial}{\partial t} h({\bf x},t) = F + \nu \nabla^2 h + \lambda
(\nabla h)^2 + \eta({\bf x},h).
\label{qkpz}
\end{equation}
Here $F$ is the driving force and $\eta({\bf x},h)$ is the quenched
disorder acting to roughen the interface; above the depinning
transition, $F \ge F_c$ leads to an interface that moves indefinitely
with a constant velocity.  Reference \cite{Amaral+Makse} finds for the
DPD class of models that the coefficient of the nonlinear term,
$\lambda$, diverges at the depinning transition \cite{qew}.

  Because of the presence of a diverging nonlinear coefficient in
(\ref{qkpz}), the application of the functional renormalization group
to the calculation of the exponents has not been possible
\cite{review}.  For this reason, the only existing estimates of the
exponents are from simulations, or from the mapping of the scaling
properties of the {\it pinned interface\/} to DP $(d = 1)$ or DS $(d >
1)$ \cite{Havlin,dpd,Buldyrev}.  Since no mapping of the {\it
dynamics\/} of Eq. (\ref{qkpz}) or of the DPD models has been found,
no theoretical estimates for the exponents characterizing the {\it
dynamics\/} of the roughening process has been made.  One such
exponent is the dynamical exponent $z$, which characterizes the
lateral propagation of the interface perturbations, as is defined by
\cite{sod},
\begin{equation}
t \sim r_{\parallel}^z.
\label{scl1}
\end{equation}
where $t$ is the time needed for a perturbation to spread over a
longitudinal distance $r_{\parallel}$.

  In this Letter we argue that $z$ for the DPD universality class can
be identified in $(d+1)$ dimensions with the percolation exponent
$d_{\rm min}$ for the shortest path for {\it isotropic\/} percolation
in $d$ dimensions \cite{perc}.  We support this relation by numerical
calculations of both $z$ and $d_{\rm min}$, up to dimension $(6+1)$.
Our work implies an upper critical dimension for the {\it dynamics\/},
$d_c + 1 = 7$, above which the mean field result, $z = 2$, becomes
exact.

  In the DPD model the growth occurs on a discrete lattice and the
disorder is modeled by considering that each cell has a probability $p$
of being ``blocked'' \cite{Havlin,dpd,Buldyrev}.  Since the model was
developed to study imbibition, we will refer to the growing, invading,
region as ``wet'', and to the remaining region as ``dry''.  At time
$t=0$, we wet all cells at the bottom of the lattice.  Then, at each
time step, we wet all dry {\it unblocked cells\/} that are
nearest-neighbors to a wet cell.  To retain a single-valued interface,
we impose the auxiliary rule that all dry {\it blocked cells\/} below a
wet cell become wet as well.  These cells we call ``eroded blocked
cells'', and this procedure is referred to as {\it erosion of overhangs}
\cite{Havlin,dpd,Buldyrev}.  If the concentration of blocked cells is
small, the interface propagates forever. As the concentration of blocked
cells increases, large portions of the interface become pinned by
fragments of DP clusters in $d=1$ or by fragments of DS for $d>1$. The
characteristic longitudinal dimension of these fragments, $\xi_\|$,
diverges as $p$ approaches the critical threshold $p_c$. When $\xi_\|$
becomes comparable with the system size $L$, the interface eventually
becomes completely pinned by a spanning DP path or DS.  Just below
$p_c$, almost all of the interface is pinned except for a few unblocked
points which move along the interface creating new sites for growth. We
address the behavior of the system only in its critical state, i.e. on
length scales smaller than $\xi_{\parallel}$, where $\xi_{\parallel}
\sim (p_c-p)^{-\nu_{\parallel}}$.

  In order to find $z$, we study how perturbations caused by a single
unblocked cell propagate over the interface. At each time step a certain
set of cells become wet by incoming fluid, caused by a single unblocked
cell at time $t=0$. In analogy with invasion percolation, we call this
set of cells the {\it percolation shell}. For each time step $t$, we
compute the average radius of gyration of the percolation shell $r(t)$.
Since $t$ is the time needed for a perturbation to spread over a
distance $r_{\parallel}$, $t$ obeys Eq. (\ref{scl1}).

  For the case $d=1$, all shells are confined between the old directed
path that spans the system at $t=0$ and a new pinning path that will
block the growth after some time. The region between these two paths
is effectively one-dimensional, since the vertical distance between
them scales as the perpendicular correlation length of DP,
$~\xi_{\perp}\sim (p_c-p)^{-\nu_{\perp}}$. Hence $\nu_\perp <
\nu_\parallel$ implies $\xi_{\perp} / \xi_{\parallel} \rightarrow 0$
as $p \rightarrow p_c$. For any cell on the interface that becomes wet
at time $t$, one can find the cell from which it was invaded at the
previous time step, and recreate the sequence of invasion events that
leads from the initial cell to any given cell on the interface
(Fig.\ref{tx1}). The trajectory of this sequence follows the upper
pinning path and is effectively one dimensional. Its length $\ell$
scales as its average end-to-end distance $r_{\parallel}$. On the
other hand, $\ell$ is equal to the time $t$ needed to reach the end of
the path.  Hence $t \sim r_{\parallel}$ and we conclude from
(\ref{scl1}) that $z=1$.  This conclusion is supported by our
simulations (Table \ref{tab1}).

  For the case $d>1$, we must consider the region bounded by two
self-affine, single-valued, DS (Fig.\ref{tx2}) \cite{Buldyrev}. This
region is effectively $d$-dimensional, since $\xi_{\perp} /
\xi_{\parallel} \rightarrow 0$. Hence, the shortest path leading from
the initial point to any point of this region is effectively confined
to a $d$-dimensional horizontal hyperplane. This shortest path has to
avoid blocked cells in this hyperplane, as does the shortest path of
isotropic percolation.  For isotropic percolation it is known that the
length of the shortest path $\ell$ scales with the Euclidean end-to-end
distance $r$ as $\ell \sim r^{d_{\rm min}}$. The similarity between
the geometrical properties of the paths in DPD and isotropic
percolation leads us to propose
\begin{equation}
z = d_{\rm min}.
\label{zdmin}
\end{equation}

  To test the argument leading to (\ref{zdmin}), we performed
simulations for both DPD and percolation for $d = 1$ to $d = 6$.  We
present our results for the exponents $z$ and $d_{\rm min}$ in Table
\ref{tab1}.

  It is well known that for isotropic percolation the upper critical
dimension is $d_c = 6$, i.e. for $d > d_c$ the mean field result,
$d_{\rm min} = 2$, becomes exact \cite{perc}.  This suggests an upper
critical dimension, $d_c + 1 = 7$, for the {\it dynamics\/} of the DPD
models which are in the universality class of Eq. (\ref{qkpz}), and
that $z = 2$ for $d + 1 \ge 7$.

  Since the dynamics of Eq. (\ref{qkpz}) and the models in the DPD
universality class are connected to isotropic percolation, while the
static properties are mapped to DP or DS, it is possible that the
upper critical dimension determined in this study {\it may be valid
only for the dynamics}.  In fact, it is possible that $d_c$ for the
static properties may not exist. Suppose, e.g., that a one-dimensional
object, such as a self-avoiding walk, is embedded in a $d$-dimensional
space.  We expect that as $d$ is increased the interactions between the
different parts of the object decrease.  At a certain $d = d_c$, these
interactions can be neglected, and the exponents become those of the
ideal non-interacting case.  In contrast, when the dimension of the
object is not fixed but increases with $d$, as in the case of DS in
which the object is one dimension smaller than the space, we expect to
move away from the non-interacting limit.  In fact, the analytical
solution of the DPD model in the Cayley tree suggests that the upper
critical dimension for the {\it statics\/} might be $\infty$
\cite{Buldyrev}.

  In summary, we present an argument that identifies the dynamical
exponent $z$ for the DPD universality class with the fractal dimension
of the shortest path in isotropic percolation, $d_{\rm min}$.  This
result leads us to identify the dimension $(6+1)$ as the upper
critical dimension for the dynamics of the DPD universality class.

        We thank A.-L. Barab\'asi, R. Cuerno, K.~B. Lauritsen, H.
Makse, and R. Sadr-Lahijany for valuable discussions.  LANA
acknowledges a scholarship from Junta Nacional de Investiga\c c\~ao
Cient\'{\i}fica e Tecnol\'ogica, and SH acknowledges partial support
from the Bi National US-Israel Foundation and from the Minerva Center
for the Physics of Mesoscopics, Fractals and Neural Networks. The
Center for Polymer Studies is supported by the National Science
Foundation.

\begin{figure}
\narrowtext
\caption{  Illustration of the dynamics of the DPD model for $(1+1)$
dimensions. (a) {\it Schematic representation \/} of a region defined
by two pinning paths.  The heavy circle indicates the origin for the
invasion, the thin arcs represent the positions of the invading front
at successive times, and the dashed line represents schematically the
path for the invasion.  (b) {\it Simulation \/} results for invasion
after $2^{10}$ time steps starting from a single cell near the center.
We show the invaded region at a sequence of times which are multiples
of $128$.  Regions invaded at later times are displayed in darker
shades of gray.  The path from the origin to the latest invaded point
is shown in black.  Although this path displays some fluctuations in
the vertical direction, they can be disregarded since $\nu_{\parallel}
> \nu_{\perp}$, so as $p \rightarrow p_c$, $\xi_{\perp} /
\xi_{\parallel} \rightarrow 0$.  Thus the distance propagated by the
invading front is proportional to time.  Since $t_{\times} \sim \ell$,
we can conclude that $z = d_{\rm min} = 1$.  }
\label{tx1}
\end{figure}

\begin{figure}
\narrowtext
\caption{  Illustration of the dynamics of the DPD model for $(2+1)$
dimensions. (a) {\it Schematic representation \/} of the $xy$
projection of the region defined by two pinning self-affine DS.  The
heavy circle indicates the origin for the invasion, the thin arcs
represent the $xy$ projections of the invading front at successive
times, and the dashed line represents schematically the path for the
invasion.  (b) {\it Simulation \/} results for invasion after $2^{10}$
time steps starting from a single cell located to the left of the
center.  We show the $xy$ projection of the invaded region at a
sequence of times which are multiples of $128$.  Regions invaded at
later times are displayed in darker shades of gray.  It is visually
apparent that it takes a long time to invade some regions close to the
origin because the path to that position (shown in black) appears to
be a fractal curve of dimension greater than one.  The fluctuations in
the vertical direction can be disregarded since we know that
$\xi_{\perp} / \xi_{\parallel} \rightarrow 0$.  We find that the path
can be identified with the shortest path (the ``chemical distance'')
of isotropic percolation, and that its length scales with the linear
distance $r$ to the point as $r^{d_{\rm min}}$.  }
\label{tx2}
\end{figure}

\begin{figure}
\narrowtext
\caption{ (a) Scaling with time of the horizontal length of a DPD
cluster in $(d+1)$ dimensions grown from a single cell.  Shown is a
double logarithmic plot of time $t$ as a function of $r_{\parallel}$,
which is the average of the parallel components of the radius of
gyration of the shell.  The asymptotic slope is $z$.  (b) Double
logarithmic plot of the shortest path $\ell$ in isotropic percolation
as a function of the Euclidean distance $r$.  The asymptotic slope is
$d_{\rm min}$.  Note that, after some transient behavior, a transition
to a power law scaling occurs.  For higher dimensions, the power law
scaling is affected by finite-size effects for larger times.  }
\label{zscal}
\end{figure}

\end{multicols}

\begin{table}
\narrowtext
\caption{ Dynamical exponent $z$ for the DPD model in $(d+1)$
dimensions and the shortest path exponent $d_{\rm min}$ for isotropic
percolation for a $d$-dimensional cubic lattice of $L^d$ sites.  The
results indicated by an asterisk are exact, while the remaining values
were calculated in our simulations by the study of the consecutive
slopes of the linear regime in Fig. 3.  At the critical dimension $d_c
= 6$, one should not expect to find the exact result $d_{\rm min}
= 2$ because logarithmic corrections are generally present.  The
system sizes used in the simulations range from $L = 4096$, for $d =
2$, to $L = 16$, for $d = 6$.  }
\begin{tabular}{c|cc|cc}
  & \multicolumn{2}{c}{DPD} &  \multicolumn{2}{c}{Percolation} \\
$~d~$      & $p_c$		& $z$            & $p_c$
	& $d_{\rm min}$	\\
\tableline
$~1~$      & $0.4698\pm 0.0002$	& $1.01\pm 0.02$ & $1^{ *}$
	& $1^{ *}$        \\
$~2~$      & $0.7425\pm 0.0002$	& $1.15\pm 0.05$ & $0.5927\pm 0.0002$
	& $1.13\pm 0.03$  \\
$~3~$      & $0.8425\pm 0.0002$	& $1.36\pm 0.05$ & $0.3116\pm 0.0002$
	& $1.38\pm 0.02$  \\
$~4~$      & $0.890\pm 0.002$	& $1.58\pm 0.05$ & $0.197\pm 0.002$
	& $1.53\pm 0.05$  \\
$~5~$      & $0.917\pm 0.003$	& $1.7\pm 0.1$   & $0.141\pm 0.002$
	& $1.7\pm 0.1$   \\
$~6~$      & $0.931\pm 0.002$	& $1.8\pm0.2$    & $0.107\pm 0.002$
	& $1.8\pm 0.2$   \\
\end{tabular}
\label{tab1}
\end{table}


\begin{references}
\bibitem{alpha}
The problem of surface roughening in the presence of quenched disorder
can be characterized by three independent exponents: $z$, $\alpha$,
and the correlation length exponent $\nu$ that characterizes the
scaling with $(p_c -p)$ of the linear size of pinned regions near the
depinning transition.  For the models in the DPD universality class,
the static exponents, $\alpha$ and $\nu$, can be determined as
functions of the exponents, $\nu_{\parallel}$ and $\nu_{\perp}$,
characterizing the correlation lengths $\xi_{\parallel}$ and
$\xi_{\perp}$ of DP clusters ($d = 1$) \cite{Havlin,dpd} and DS ($d >
1$) \cite{Buldyrev}.  For this universality class, it can be shown
that $\nu = \nu_{\parallel}$ and $\alpha = \nu_{\perp} /
\nu_{\parallel}$.
\bibitem{qew}
For a number of different models, belonging to a different
universality class, it was found that either $\lambda = 0$ or $\lambda
\rightarrow 0$ at the depinning transition \cite{Amaral+Makse}.  So,
near the depinning transition, they can be described by Eq. (1) with
$\lambda = 0$.  This equation has been studied by means of the
functional renormalization group \cite{review}.
\end{references}
\end{document}